\documentstyle{b98proc}

\def\Journal#1#2#3#4{{#1} {\bf #2}, #3 (#4)}

\def\NPA{{\em Nucl. Phys.} A}

\def\PLB{{\em Phys. Lett.} B}

\def\PRD{{\em Phys. Rev.} D}
\def\PRC{{\em Phys. Rev.} C}

\begin{document}

\rightline {Tel Aviv U. preprint TAUP-2568-99}
\begin{center}
\noindent {\large \bf
INELASTIC ELECTRON-PION SCATTERING\\ at FNAL (SELEX)}

\vspace{6pt}
%


The SELEX Collaboration \\ 
M.A.~Moinester$^{12}$,
A.~Ocherashvili$^{12}$,
V.~Steiner$^{12}$,
N.~Akchurin$^{16}$,
V.A.~Andreev$^{11}$,
A.G.~Atamantchouk$^{11}$,
M.~Aykac$^{16}$,
M.Y.~Balatz$^{8}$,
N.F.~Bondar$^{11}$,
A.~Bravar$^{20}$
% ...

% ...
$^8$Institute of Theoretical and Experimental Physics, Moscow, Russia\\
% ...
$^{11}$Petersburg Nuclear Physics Institute, St. Petersburg, Russia\\
$^{12}$Tel Aviv University, 69978 Ramat Aviv, Israel\\
% ...
$^{16}$University of Iowa, Iowa City, IA 52242, U.S.A.\\
% ...
$^{20}$University of Trieste and INFN, Trieste, Italy\\

\vspace{6pt}
\noindent{This is a contribution for the Baryons '98 conference,
Univ. of Bonn, Sept. 1998.}\\[1mm]

\end{center}

\vspace{6pt}

\abstracts{
We describe the analysis status of SELEX
electron-pion inelastic  $\pi e \rightarrow \pi' e' \gamma$  and $\pi e
\rightarrow \pi' e' \pi^0$ reaction data. }



\section{Introduction}

The SELEX experiment (E781) at Fermilab \cite{selex} focused on charm
hadroproduction \cite{ichep} at large X$_F$, using 590 GeV/c $\pi^-$ and
$\Sigma^-$ and proton beams. We took data simultaneously with the same
beams and electron targets (atomic electrons in nuclear target) for
elastic \cite{ie} and inelastic hadron-electron scattering. Here, we
describe the analysis status of electron-pion inelastic scattering $\pi e
\rightarrow \pi' e' \gamma$ and $\pi e \rightarrow \pi' e' \pi^0$ reaction
data. We will discuss these reactions in terms of the kinematic variables
defined in Fig. 1a.  The data give information on reactions that were
never previously measured:  (1) $\pi e \rightarrow \rho e'$ scattering for
a determination of the $\rho \rightarrow \pi \gamma$ radiative width from
a measurement of the transition form factor (FF) near zero momentum
transfer, (2) $ \pi e \rightarrow e' \pi' \pi^0$ scattering near threshold
for a determination of the chiral anomaly transition FF and the $\gamma
\rightarrow 3 \pi$ F$_{3\pi}$ chiral anomaly amplitude, and (3) $\pi e
\rightarrow \pi' e' \gamma $ scattering, in which a virtual photon from
the electron's Coulomb field is Compton scattered on the pion, for a
determination of the never previously measured generalized pion
polarizabilities. 

\section{The SELEX Setup}

The SELEX three-stage magnetic spectrometer allowed us to measure the complete
kinematics of the reaction. Scattering angles and momenta of
both the hadron and the electron were measured with high precision using
silicon microstrip detectors located before and after the targets, and at
small angles after the SELEX magnets. There are also proportional wire and
drift chambers for measuring charged tracks. The tracking planes are organized
in spectrometers, interspaced with dipole magnets for momentum analysis. There
are three lead glass calorimeters for photon detection and electron
identification after each magnet. There is a transition radiation detector
(TRD) that tags the type of beam particle, and a second TRD after the second
magnet for electron identification. There is a ring imaging Cherenkov
counter
(RICH) for particle identification. For the trigger, we use an ensemble of
fast scintillation detectors for beam definition, two scintillation
interaction
counters which measure an energy loss appropriate to two final state charged
particles, and a scintillation hodoscope downstream after the SELEX
magnets to
also identify two negative charged particles.

\section{Physics Topics\label{sec:physics}}

\subsection{Excitation of the $\rho$}

Data analysis is in progress for 
the reaction $ \pi e \rightarrow  e'
\pi' \pi^o$,
to identify the $\pi e \rightarrow \rho e'$ chanel. Such data 
will allow a determination of the $\rho \rightarrow \pi \gamma$ radiative
width from a measure of the transition FF (near zero momentum transfer). For
this study, the two detected $\gamma$'s must have a $\pi^0$ invariant mass,
and the $\pi\pi^0$ system must have the invariant mass of the $\rho$.

The $\pi^-$ has J$^{\pi}=0^-$ and the $\rho^-$ has J$^{\pi}=1^-$. A spin-flip
M1 transition is required. The transition FF is given (schematically) by the
overlap integral:\\ $F(q^2)=\int \Psi[\rho(\vec{r})]^* \Psi[\pi(\vec{r})]~
\exp(i\vec{q}\vec{r})~[\sigma Y1(\vec{r})]~d\vec{r},$ \noindent where
$\Psi[\rho(\vec{r})]$ and $\Psi[\pi(\vec{r})]$ are $\rho$ and $\pi$
wave functions, $\sigma$ is the spin flip operator, and $\vec{q}$ is the
momentum transfer. The FF probes our understanding of the pion and rho wave
functions \cite {los,ia}. The $\rho$ radiative width ($\Gamma(\rho \rightarrow
\pi \gamma \approx 70~ KeV$) is fixed by the value of the FF at $q^2=0$. There
is a known  relationship of the radiative width to the FF \cite {ia}. Our
studies of the $\rho$ channel will be also valuable for understanding the
chiral anomaly reaction $ \pi e \rightarrow  e' \pi' \pi^0$.

\subsection{Chiral Anomaly Transition}

For the $\gamma$-$\pi$ interaction, the O(p$^4$) chiral lagrangian includes
Wess-Zumino-Witten (WZW) terms \cite{hols2,ca,bij3}, which lead to an abnormal
intrinsic parity (chiral anomaly) term in the divergence equations of the
currents. Data for $e \pi \rightarrow e' \pi' \pi^0$, where the $\pi\pi^0$
system has invariant mass lower than the $\rho$, near threshold, should
determine the chiral anomaly transition FF $\pi  \gamma  \rightarrow \pi' 
\pi^0$, and $\gamma \rightarrow  3 \pi$ F$_{3\pi}$ amplitude at threshold. The
O(p$^4$) prediction \cite{hols2,ca} is $F_{3\pi}$ = 9.7 GeV$^{-3}$. Holstein
\cite{hols2} gives the transition FF as F~=~F$_{3\pi}$ $\times$ f, where f is
a known function of the kinematic variables. The absolute cross section data
for $ \pi e \rightarrow e' \pi' \pi^0$ should determine F$^2_{3\pi}$, and the
kinematic dependence should determine the shape f$^2$.

   Antipov et al. \cite{anti2} measured F$_{3\pi}$ with 40 GeV pions. Their
study involved pion production by a pion in the nuclear Coulomb field via the
Primakoff reaction: $\pi^- Z \rightarrow {\pi^-}'  \pi^0  Z'.$ The Antipov
et al. \cite{anti2} experiment (with roughly 200 events) yielded
F$_{3\pi}=12.9 \pm 0.9 (stat) \pm 0.5 (sys) ~GeV^{-3}$. A reanalysis by
Holstein \cite{hols2} gave F$_{3\pi}$ lower by 1 GeV$^{-3}$. Bijnens et al.
\cite{bij3} studied higher order $\chi$PT corrections. For F$_{3\pi}$, they
increase the theoretical prediction by around 1 GeV$^{-3}$. The prediction at
O(p$^6$) is then $F_{3\pi} \sim  10.7$, closer to the data. The limited
accuracy of the existing data, together with the new calculations of Bijnens
et al., motivate an improved and more precise experiment.

\subsection{Pion Generalized Polarizabilities\label{sec:gen_pol}}

The virtual Compton scattering (VCS) reaction $\pi e \rightarrow \pi' e'
\gamma$ is sensitive to the generalized pion electric and magnetic
polarizabilities $\bar{\alpha}_{\pi}$(q) and $\bar{\beta}_{\pi}$(q), which
depend on momentum transfer (q) to the electron \cite {dd,guic}. At zero
momentum transfer, these reduce to the usual Compton polarizabilities \cite
{babu2}. The VCS process and planned proton VCS experiments at
electron accelerators have been discussed extensively \cite {ndh}. S. Scherer
et al. \cite {ss} are calculating pion VCS.

For pion VCS, the Bethe-Heitler (BH) amplitude ($\gamma$ from electron)
dominates over the Compton (C) amplitude. But the Compton amplitude should be
relatively more enhanced compared to BH for events in which the angle between
$\gamma$ and electron is large. The complete transition amplitude is BH+C,
with the BH amplitude much larger than the C amplitude. The cross section
depends on BH$^2$ $\pm$ 2 (BH) (C) +C$^2$, where the sign changes between
positive and negative pion beams \cite {ss}. 
We will compare our $\pi^-$ VCS data to the theoretical calculations, by
incorporating the theory into the SELEX GEANT simulation package \cite
{ge781}. The object is to fit the data to obtain the generalized
polarizabilities with their momentum transfer dependences, and also the q=0
Compton limit.

\section{Data Analysis Status}

\subsection{Excitation of the $\rho$}
In Fig. 1b, we show a reconstruction of the $\pi+2\gamma$
invariant mass, when the $\gamma\gamma$ invariant mass is around the $\pi^o$
mass. 
Only events with two detected $\gamma$-rays above 1 GeV were considered.
Fig. 1b is a raw analysis spectrum, not with the final
analysis package, not with all possible kinematic cuts, and not acceptance
corrected. Such is the case also for other preliminary spectra shown in this
paper. But we can already clearly see signs of a $\rho$ resonance peak close
to the expected mass. Data analysis methods are
being improved, and should reduce backgrounds. With lower backgrounds
and acceptance corrections, we may look at the lower mass chiral
anomaly region. We will again compare theory and data with the help of the
SELEX GEANT package \cite {ge781}. The yield falls with q$^2$,
as shown in Fig. 1c. 

\subsection{Generalized Polarizabilities}

In $\pi e \rightarrow~ \pi \gamma e$, a virtual photon from the electron's
Coulomb field is Compton scattered on the pion.  Preliminary data from the
1-$\gamma$ channel are shown in Fig. 1d, for events with only one
detected $\gamma$-ray above 1 GeV. The data are due to 1-$\gamma$ physics
events, and part is due to "1-$\gamma$" background from the $\pi e \rightarrow
\pi \pi^0 e$ channels for events in which one $\gamma$ from the $\pi^0$ decay
does not satisfy all our detection criteria. 
We are making first estimates of
this background by taking our $\pi e \rightarrow \rho e'$ production rate,
building an event generator for this process, and using the SELEX GEANT
simulation package. Our aim is to subtract the estimated background from the
measured 1-$\gamma$ spectra. Our simulation work is in progress, so that the
resulting background subtracted spectra are not yet available. 
The S$_1$ dependence of the data is shown in Fig. 1d. 
The data peaks at the lowest S$_1$ bin, 
as expected for the Bethe-Heitler process.

\section{Conclusions}

We describe first electron-pion inelastic data via the SELEX experiment. We do
not give any final results for the pion-electron inelastic channels. Our aim
here is rather to show work in progress, and to help generate needed
theoretical support.

\section{Acknowledgements}

The Tel Aviv U. group acknowledges support by the U.S.-Israel Binational
Science Foundation (BSF) and the Israel Science Foundation founded by the
Israel Academy of Sciences and Humanities.


\begin{figure}[H]
  \begin{center}
    \leavevmode
    \centerline{\epsfig{file=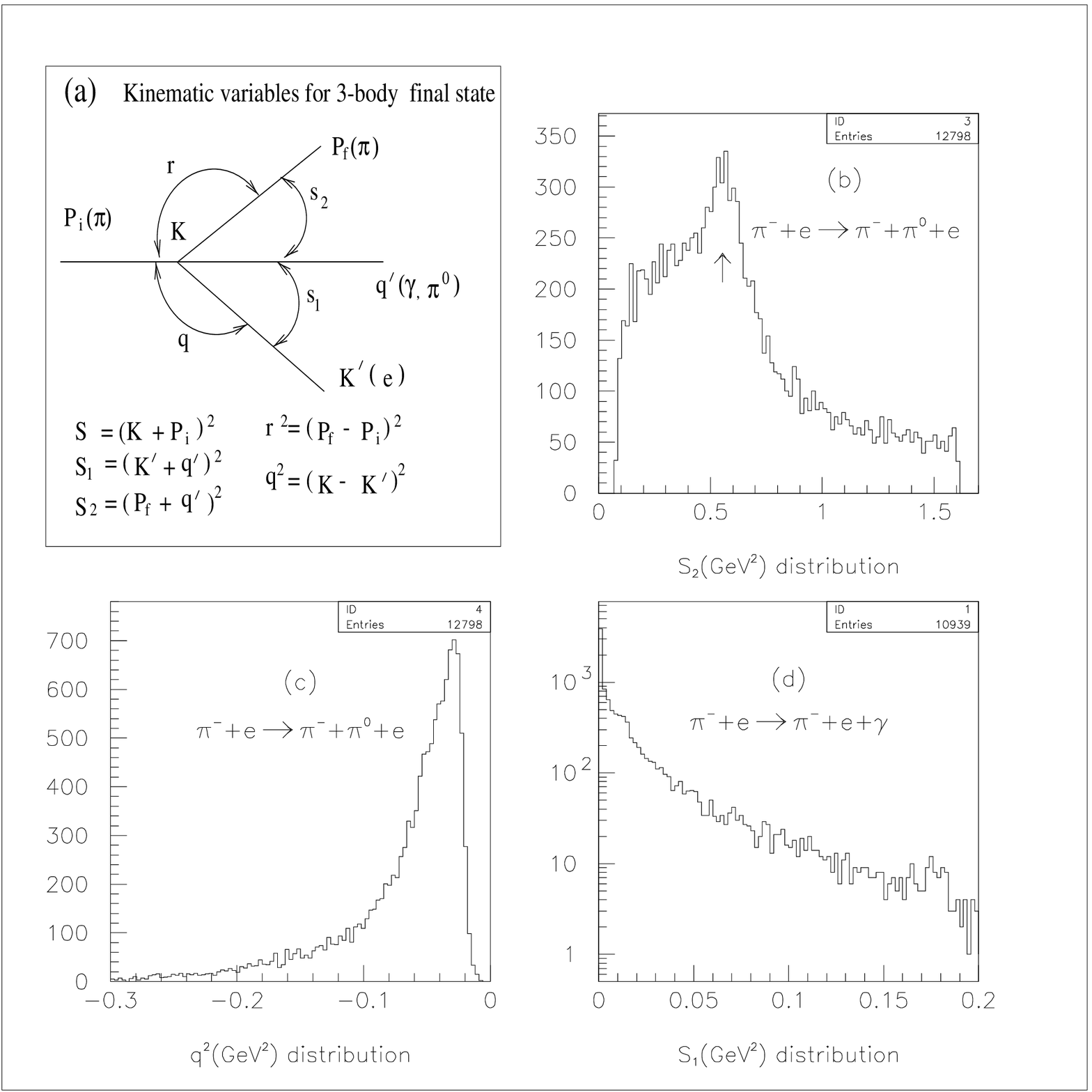,width=15cm,height=15cm}}
    \caption{(a) Kinematic variables for reaction 
$\pi^-+e\rightarrow\pi^-+e+\gamma$ and
$\pi^-+e\rightarrow\pi^-+e+\pi^0$. $P_i$ is the 4-momentum of the incoming
$\pi^-$ beam, $K$ is the 4-momentum of the $electron$ target, $P_f$ is the
4-momentum of the outgoing $\pi^-$, $q'$ of the outgoing $\gamma$ in case
of $\pi^-+e\rightarrow\pi^-+e+\gamma$ or $\pi^0$ in case of
$\pi^-+e\rightarrow\pi^-+e+\pi^0$, $K'$ is the 4-momentum of the outgoing
$electron$.~~SELEX preliminary data (b,~c): Distributions of events 
versus S$_2$ and q$^2$ for the reaction
$\pi^-+e\rightarrow\pi^-+e+\pi^0$. The arrow in (b) gives the expected
position of the $\rho$.~~
SELEX preliminary data (d): Distributions of events 
versus S$_1$ for the reaction
$\pi^-+e\rightarrow\pi^-+e+\gamma$.}
    \label{fig:kin}
  \end{center}
\end{figure}

\end{document}